\documentclass[pra,twocolumn,superscriptaddress,aps,floatfix,longbibliography]{revtex4-2}

\usepackage{hyperref}
\usepackage{amsmath}
\usepackage{bm}
\usepackage{graphicx}
\usepackage[ansinew]{inputenc}
\usepackage{array}
\usepackage{color}
\usepackage{amsxtra} 
\usepackage{amstext}
\usepackage{amssymb}
\usepackage{latexsym}
\usepackage{dsfont}
\usepackage{braket}
\usepackage[caption=false]{subfig}
\usepackage[makeroom]{cancel}
\usepackage{lineno}
\usepackage{physics}
\usepackage{orcidlink}

\usepackage{dcolumn}
\usepackage{bm}
\usepackage{bbm}
\usepackage{braket}
\usepackage{mathrsfs}
\usepackage{amstext}
\usepackage{euscript}

\usepackage{hyperref}

\usepackage{txfonts}

\usepackage{xcolor}
\definecolor{dgreen}{rgb}{0.0,0.6,0.0}
\definecolor{pink}{rgb}{1,0,0.9}

\usepackage[normalem]{ulem}

\newcommand{\ms}[1]{\mbox{\scriptsize #1}}
\newcommand{\msi}[1]{\mbox{\scriptsize \textit{#1}}}

\newcommand{\cref}[1]{Ref.\,\cite{#1}}

\begin{document}

\title{Hybrid non-degenerate parametric amplifier for a microwave cavity mode and an NV ensemble}

\author{Roman Ovsiannikov\orcidlink{0009-0003-9558-4132}}
\email{roman.ovsiannikov@kipt.kharkov.ua}
\affiliation{Akhiezer Institute for Theoretical Physics, NSC KIPT, Akademichna 1, 61108 Kharkiv, Ukraine}

\author{Kurt Jacobs\orcidlink{0000-0003-0828-6421}}
\email{dr.kurt.jacobs@gmail.com}
\affiliation{Advanced Photonics and Electronics Division, U.S. Army DEVCOM Army Research Laboratory, Adelphi, Maryland 20783, USA} 
\affiliation{Department of Physics, University of Massachusetts at Boston, Boston, Massachusetts 02125, USA}

\author{Andrii G. Sotnikov\orcidlink{0000-0002-3632-4790}}
\email{a\_sotnikov@kipt.kharkov.ua}
\affiliation{Akhiezer Institute for Theoretical Physics, NSC KIPT, Akademichna 1, 61108 Kharkiv, Ukraine}
\affiliation{Education and Research Institute ``School of Physics and Technology'', Karazin Kharkiv National University, Svobody Square 4, 61022 Kharkiv, Ukraine}

\author{Matthew E. Trusheim} 
\email{mtrush@mit.edu}
\affiliation{Advanced Photonics and Electronics Division, U.S. Army DEVCOM Army Research Laboratory, Adelphi, Maryland 20783, USA}  
\affiliation{Department of Electrical Engineering and Computer Science, Massachusetts Institute of Technology, 77 Massachusetts Avenue, Cambridge, Massachusetts 02139, USA}

\author{Denys I. Bondar\orcidlink{0000-0002-3626-4804}}\email{dbondar@tulane.edu}
\affiliation{Department of Physics and Engineering Physics, Tulane University,  New Orleans, Louisiana 70118, USA}

\date{\today}

\begin{abstract} 

We analyze a theoretical proposal for a hybrid non-degenerate parametric amplifier in which the signal and idler modes are, respectively, a microwave cavity mode and the collective mode of a spin ensemble, such as an ensemble of nitrogen-vacancy centers in diamond. We show that within the linearized collective-spin model, modulation of the spin transition frequency near the sum of the cavity and spin frequencies generates an effective parametric amplification process. This mechanism can generate two-mode squeezing, and we present a control protocol that converts this two-mode squeezing into single-mode squeezing of the cavity and spin degrees of freedom. The effect arises from the interplay between longitudinal spin-frequency modulation and spin-cavity coupling. For representative cavity parameters, our simulations predict approximately \(4~\mbox{dB}\) of microwave signal amplification in \(2~\mbox{$\mu$s}\) for a modulation amplitude \(\Lambda/(2\pi)=200~\mbox{MHz}\), in which the bandwidth is the inverse of the amplification time.  At \(10~\mbox{mK}\) with cavity and spin quality factors \(Q=5\times10^4\) and a modulation amplitude of \(\Lambda/(2\pi)=500~\mbox{MHz}\) the model gives approximately \(3~\mbox{dB}\) of squeezing in terms of the quadrature standard deviation. We also discuss the key challenges in realizing this parametric amplifier using nitrogen-vacancy centers.
\end{abstract}

\maketitle

\section{Introduction}

The creation of non-classical squeezed states, in particular using highly-sensitive spin ensembles,~\cite{spin_squeezing_atoms,NV_squeezing_recent,NV_cavity_QED_review} can potentially power prospective quantum technologies in sensing~\cite{squeezing_sensing_review} and communication~\cite{squeezing_communication_review}. A standard approach  to squeezing is phase-sensitive amplification, such as the optical parametric oscillators used in the LIGO experiment~\cite{LIGO2023}. Amplification in the microwave frequency range of electron spin degrees of freedom is ubiquitous, with numerous applications in the classical ~\cite{HEMT_review,Coaker_Challis_2008} and quantum domains~\cite{JPA_review,JPA_squeezing_review}. Direct production of a squeezed spin state, however, is complicated by the fact that unlike parametric driving of an oscillator, the parametric driving of an isolated spin ensemble will not generate squeezing or amplification of the ensemble state. Previous studies have explored the production of squeezing by using an additional $2^{\msi{nd}}$-order nonlinear medium to parametrically drive a microwave mode~\cite{qin2018exponential, leroux2018enhancing, zhang2021heisenberg}, producing squeezing which can then be transferred to the spins via spin-cavity coupling.  Here we theoretically show that this additional nonlinear element is not required:
modulation of the spin transition frequency produces a non-degenerate parametric amplification process for the microwave mode and collective spin mode, with the resulting amplification and two-mode squeezing. The two-mode squeezing can be transformed into single-mode squeezing via the spin-cavity coupling.
 
Taking optically polarized spin states in nitrogen-vacancy (NV) color centers as a prototypical platform, we use experimentally motivated parameters to estimate the performance of the proposed scheme. The model predicts that parametric driving can provide over \(30~\mbox{dB}\) gain with a bandwidth of order \(0.5~\mbox{MHz}\), and with a noise temperature comparable to recent solid-state maser demonstrations~\cite{day2024maseramp}. The same mechanism would also amplify signals stored in the spin ensemble, suggesting a possible route to room-temperature spin-ensemble signal processing. Optical pumping can make the effective spin noise much lower than the microwave thermal occupation at room temperature, which is the key advantage of using a polarized spin ensemble as the idler mode.

To realize squeezing with the parametric amplifier requires cryogenic temperatures (\mbox{$\lesssim 100$ mK}). Further, any heating or quasi-particle production caused by the microwave drive would have to be sufficiently low. Assuming this can be achieved, the model shows that parametric driving of the spins generates two-mode squeezing similar to that produced by a hybrid non-degenerate parametric amplifier for the cavity mode and the spin ensemble. Without cavity loss, the rate of squeezing is about $\mbox{5 dB/$\mu$s}$ at a modulation amplitude $\Lambda/(2\pi)=500~\mbox{MHz}$. At \(10~\mbox{mK}\) with cavity and spin quality factors \(5\times 10^5\), the model gives \(3~\mbox{dB}\) of squeezing. For quality factors of \(1.25\times 10^4\), this decreases to \(1.8~\mbox{dB}\) of squeezing. While the driving of the spins generates two-mode squeezing between the cavity mode and the spin ensemble, we show that the detuning of the spins with respect to the cavity and the collective spin/cavity interaction can be used in a time-dependent protocol to transform this two-mode squeezing into simultaneous single-mode squeezing of the cavity and the ensemble. While the performance of the single-mode microwave squeezing is not as high as that of Josephson parametric amplifiers (JPAs), the ability to squeeze the spin degree of freedom (either singly or as two-mode squeezing) is a qualitative advantage. Furthermore, the NV-cavity system can be operated at higher temperatures and potentially with higher-$Q$ systems.

The paper is laid out as follows. In the next section, we describe the physical system and the theoretical description that we use. In Sec.~\ref{covariance}, we define the single-mode and two-mode covariance matrices for the quadratures of the cavity mode and the spin ensemble. The latter can be treated as a harmonic oscillator, as the spins remain in their linear regime. We also define the use of decibels for amplification and squeezing. In Sec.~\ref{pert}, we show that a time-dependent perturbative analysis of the parametrically driven system can be used to gain insight into the modulation frequency required to obtain amplification and squeezing. In Sec.~\ref{nums} we present numerical results from simulations of the system, the results of our analysis of the noise temperature of the amplifier, and the control protocol that converts between the two-mode and single-mode squeezing. Section~\ref{expreal} discusses implementation requirements and feasibility estimates for a possible NV-cavity realization of the proposed hybrid parametric amplifier. Finally, in Sec.~\ref{conc} we conclude with a summary of our main findings. 

\begin{figure}[t]
    \centering
\includegraphics[width=0.9\columnwidth]{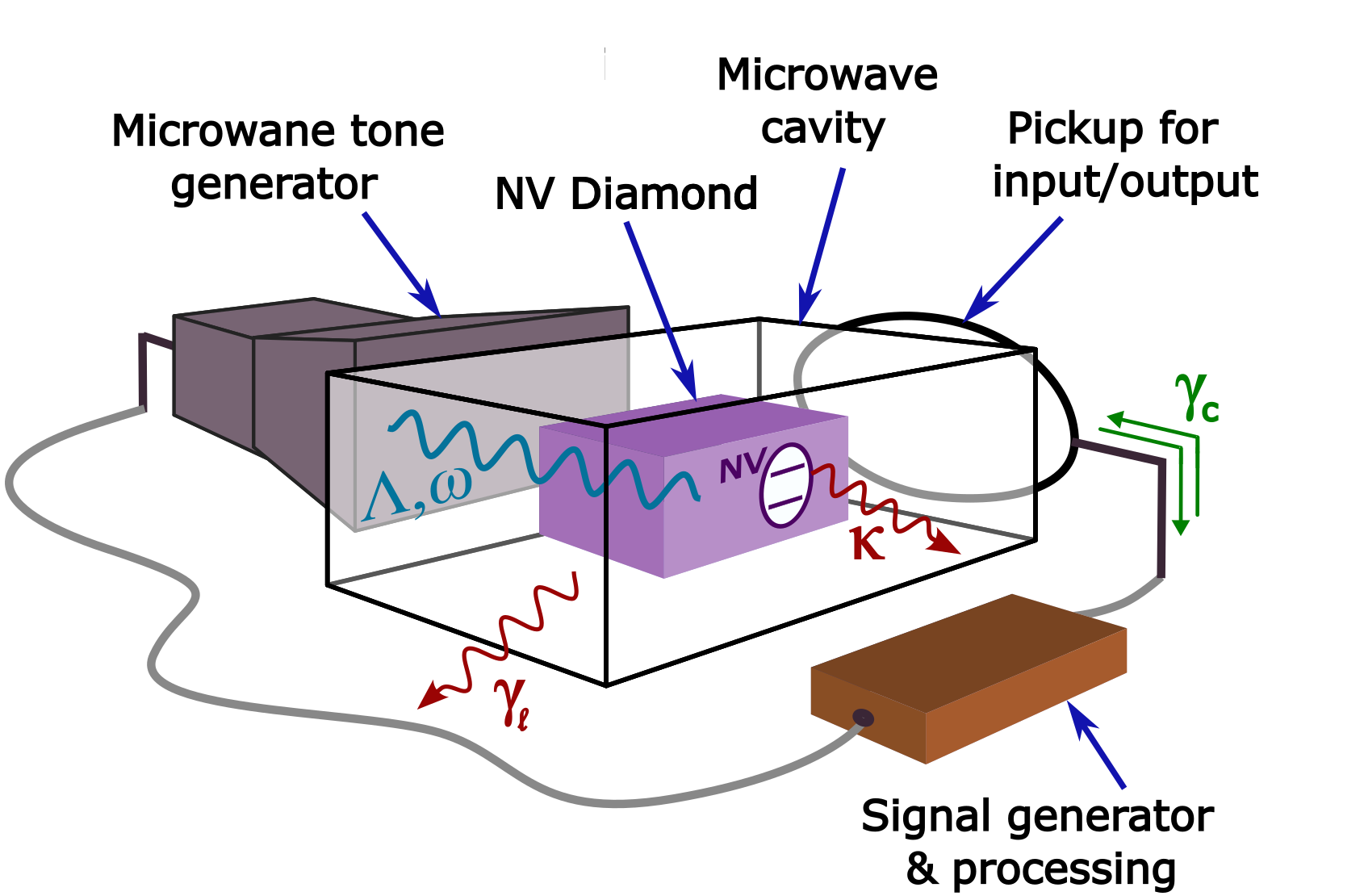} 

\caption{Schematic of the proposed NV-cavity parametric amplifier. A diamond sample containing an ensemble of NV centers is coupled to a microwave cavity mode that carries the signal. A pump microwave field, supplied by a second mode of the cavity or by a mode of an overlapping second cavity, modulates the NV transition frequency near the sum of the cavity and spin frequencies. The spin transition is polarized by optical pumping. Here \(\omega\) and \(\Lambda\) are the frequency and amplitude of the microwave modulation, \(\gamma_{\mathrm c}\) is the cavity input-output coupling rate, \(\gamma_{\mathrm l}\) is the cavity internal loss rate, and \(\kappa\) is the effective linewidth of the collective spin mode used in the reduced model. The spin-cavity detuning \(\Delta\) and collective coupling rate \(g\) are not shown.}
    \label{System_diagram}
\end{figure}


\section{The Coupled NV/Cavity system}
 
We consider a single mode of a microwave resonator interacting with an ensemble of $N$ nitrogen-vacancy centers. These NV centers are 3-level systems in the ``V" configuration with the two upper states having spin projections $m_s=\pm 1$ respectively, and in the absence of a magnetic field both transitions are at approximately $3~\mbox{GHz}$. We aim to couple the cavity mode to a single transition only. To do this, we apply a magnetic field which shifts one transition by $-2.5~\mbox{GHz}$ to $500~\mbox{MHz}$ and the other to $5.5~\mbox{GHz}$. We place our microwave cavity frequency at $1.5~\mbox{GHz}$ so that it interacts off-resonantly with the lower transition at a detuning of $- 1~\mbox{GHz}$. The higher-frequency transition is then detuned from the cavity by 4 GHz. Numerical simulations show that the effect of a transition at $4~\mbox{GHz}$ detuning modifies the quadrature variances by less than $1\%$ over 4000 periods of the drive. At this level of approximation we can thus treat the NVs as two-level systems, which we do from now on.

For compactness, throughout this paper, for any given Hamiltonian, $H$, we will define the \textit{rate Hamiltonian} as $\tilde{H} \equiv H/\hbar$, denoting it with a tilde. This definition suffices to remove many factors of $\hbar$ throughout the presentation, while preserving real units for all results. 

We will apply an oscillating magnetic drive to the NVs to modulate their transition frequency. The resulting rate Hamiltonian for the coupled cavity/NV system, depicted in Fig.~\ref{System_diagram}, can be expressed under the usual cavity-QED approximations~\cite{Tannoudji97, Agarwal12}  as
\begin{align}\label{eq:fullH}
    \tilde{H} = \omega_{\ms{c}} a^\dagger a + [\omega_{\ms{s}} + \Lambda  \sin(\omega t)]\dfrac{S_z}{2} 
    +  g (a + a^\dagger)(S_- + S_-^\dagger) . 
\end{align} 
Here $\omega_{\ms{c}}$ is the cavity mode frequency, $\omega_{\ms{s}}$ is the frequency of the NV centers in the absence of modulation, and $\Lambda$ and $\omega$ are the amplitude and frequency of the modulation, respectively. The cavity annihilation operator is $a$, while $S_-$ and $S_z$ are, respectively, the sums over the lowering operators $\sigma^{(j)}_{-}$ and the spin $z$-component operators $\sigma_z^{(j)}$ for the NV centers:
\begin{align}
    S_{-} & = \sum_{j=1}^N \sigma^{(j)}_{-} , \;\;\;\;\;
    S_z  = \sum_{j=1}^N \sigma_z^{(j)} . 
\end{align}
Here we use Pauli operators $\sigma_z^{(j)}$ with eigenvalues $\pm 1$, so that the single-spin rate Hamiltonian contains the factor $1/2$ in Eq.~(\ref{eq:fullH}). The collective angular-momentum operators used below are defined as
\begin{align}\label{eq:Jdefs}
    J_z = \frac{S_z}{2}, \qquad J_- = S_-, \qquad J_+ = S_+ .
\end{align}

The size of the Hilbert space of the coupled system grows exponentially with the number of NVs, which we will also refer to as \textit{spins}. Fortunately, since the interaction between the cavity and the spins is completely symmetric in the spins, as long as the spins begin in their ground states, the dynamics of the system can only populate completely symmetric superposition states of the spins (the Dicke states)~\cite{Dicke54}. As a result, it is only the spin space with the largest value of total angular momentum ($N/2$) that plays a role in the dynamics, and we can replace the operators $S_{-}$ and $S_z$ respectively with the lowering (ladder) operator $J_-$ and the $z$-component operator $J_z$ for this value of total angular momentum. The dimension of the accessible space for spin states is then $N+1$. The rate Hamiltonian becomes  
\begin{equation}\label{eq2}
    \tilde{H} = \omega_{\ms{c}} a^\dagger a + [\omega_{\ms{s}} + \Lambda  \sin(\omega t)]J_z + g (a + a^\dagger)(J_- + J_+) , 
\end{equation}
with $J_+ = J_-^\dagger$. We now employ the Holstein-Primakoff representation for the $N/2$ angular momentum operators~\cite{holstein1940}, which is
\begin{equation}
    J_z = b^\dagger b - N/2, \quad J_- = \sqrt{N} \sqrt{1 - \dfrac{b^\dagger b}{N}} b,
\end{equation}
where $b$ ($b^\dagger$) is the boson annihilation (creation) operator.   
Note that since we are free to add a constant to any term in the Hamiltonian, we can replace $J_z$ simply with $b^\dagger b$. When the number of excitations in the spin system remains much less than $N/2$, the spins are in a ``linear'' regime: we can make the approximation $J_- = \sqrt{N} b$ so that the spin system acts for most purposes as a harmonic oscillator. The spin system remains rather distinct from a harmonic oscillator in one respect, however; the lowering operator $b$, arising from the Holstein-Primakoff representation, is independent of the spin frequency and is thus unaffected by the frequency modulation. For a harmonic oscillator, on the other hand, since the annihilation operator contains the frequency as part of its definition, it is itself time-dependent under frequency modulation. This is the reason that harmonic oscillators are squeezed by parametric driving while spin systems are not. 

Even though the NV centers are only polarized to $80\%$ by the pump laser, the Holstein-Primakoff approximation is usually assumed to work well since experiments show linear behavior at this level of polarization. We nevertheless confirm the accuracy of this approximation using numerical simulations in Appendix~\ref{HPcheck}. 

With the spins remaining in their linear regime, we can now write the rate Hamiltonian as 
\begin{equation}
    \tilde{H}_{\rm HP} = \omega_{\ms{c}} a^\dagger a + [\omega_{\ms{s}} + \Lambda  \sin(\omega t)] b^\dagger b + g (a + a^\dagger)(b + b^\dagger).  
    \label{Ham}
\end{equation}

So far we have assumed that the spins all have a single frequency, $\omega_{\mathrm{s}}$. We must now take into account that the spins have a distribution of frequencies, usually referred to as inhomogeneous broadening. The full width at half maximum (FWHM) of this distribution is typically a few hundred kHz~\cite{Bauch20, Wang24} and is very much larger than the dephasing and damping rates of the individual spins. The effect of this frequency spread is that the spins form a thermal bath for the cavity mode. To see this, note that there is a very large number of spins ($N\sim 10^{14}$). If we divide the spin frequency distribution into thin slices, we see that each ensemble of spins within a slice acts as a linear oscillator with a given frequency. Thus the spins are equivalent to a bath of harmonic oscillators with a spectrum given by their frequency distribution. While this frequency distribution can be expected to be Gaussian, if we make the approximation that it is Lorentzian, then we can obtain a particularly simple and tractable model that describes the spins. We use the fact that coupling a system to a bath with a Lorentzian spectrum is exactly equivalent to coupling the system to a single (fictitious) harmonic oscillator, which is itself coupled to a Markovian (flat spectrum) bath with a damping rate equal to the FWHM of the Lorentzian~\cite{Imamoglu94,Garraway97,Dalton01,Fahey23}. This has been referred to as the ``pseudo-mode" description of colored baths. We thus arrive at a description of the spins as a single oscillator that is damped at a rate equal to the spin inhomogeneous linewidth. Following this terminology we will refer to the fictitious oscillator as the pseudomode.

Finally, the overall frequency of the spins (the mean of the frequency distribution) is modulated sinusoidally at frequency~$\omega$. As discussed above, this changes only the frequency of the effective harmonic oscillators describing the spins, but not the annihilation and creation operators which would be the case if the spins were actually harmonic oscillators. So in our effective description of the spin bath, the frequency of the pseudomode is changing by exactly the same amount as the frequencies of the oscillators in the Markovian bath. Under the rotating wave approximation, the sum of all the oscillator Hamiltonians (the pseudomode and the bath oscillators) commutes with the interaction between them. A direct result of this is that the time dependence of the spin frequencies has no effect on the master equation for the pseudomode: the only effect of the frequency modulation is to modulate the frequency of the pseudomode.

The intrinsic decay rate of the spins from the upper spin level to the ground state is less than 1 kHz, and the damping rate induced by the laser that polarizes the spins to the ground state is about 2 kHz~\cite{Fahey23}. Since the damping induced by the inhomogeneous broadening of the spins is much greater than this quantity, we do not need to include the spin damping in our model.


We have now arrived at our full description of the cavity mode and the frequency-modulated ensemble of spins. Including the damping of the cavity mode at rate $\gamma$, and denoting the inhomogeneous linewidth of the spins by $\kappa$, we have the following Markovian master equation for the cavity mode and the oscillator describing the spins (the psuedomode)~\cite{Jacobs14}: 
\begin{align}\label{Lindb_eq}
        \dfrac{d\rho }{dt}& = i[\rho, \tilde{H}_{\rm HP}] + \frac{\gamma}{2}(1 + n_{T}) \left( 2 a \rho a^\dagger -a^\dagger a \rho - \rho a^\dagger a \right) \notag \\
        & + \frac{\gamma}{2} n_{T} \left( 2 a^\dagger \rho a - a a^\dagger \rho - \rho a a^\dagger \right) \notag\\
        &+ \frac{\kappa}{2}(1 + n_{\mathrm{s}})\left( 2 b \rho b^\dagger -b^\dagger b \rho - \rho b^\dagger b \right) \notag \\
        &+\frac{\kappa}{2} n_{\mathrm{s}} \left( 2 b^\dagger \rho b - b b^\dagger \rho - \rho b b^\dagger \right) . 
\end{align} 
The cavity damping rate $\gamma = \gamma_{\mathrm{c}}+\gamma_{\mathrm{l}}$ in which $\gamma_{\mathrm{c}}$ is the cavity's input/output coupling rate and $\gamma_{\mathrm{l}}$ is the cavity's internal loss rate. The quantity $n_{T}$ is the average number of photons that the cavity mode would have if it were in thermal equilibrium at temperature $T$, which is the ambient temperature of the cavity. The value of $n_T$ is obtained from the corresponding Bose--Einstein distribution function, which leads to 
\begin{align}
    n_{T} = \frac{1}{\exp\left[\hbar\omega_c/(k_{\mathrm{B}}T) \right] - 1}, 
\end{align} 
in which $k_{\mathrm{B}}$ is the Boltzmann constant. The quantity $n_{\mathrm{s}}$ is the number of photons that the spin oscillator would have if it were at the temperature of the spins, $T_{\mathrm{s}}$. At $80\%$ polarization and transition frequency $500~\mbox{MHz}$ we have $n_{\mathrm{s}} = 1/8$ and $T_{\mathrm{s}} \approx 11~\mbox{mK}$.  

\section{Individual and joint quadratures}
\label{covariance}

Since we are concerned with parametric amplification and squeezing of both modes in the two-mode system, we are interested in the respective 2$\times$2 covariance matrices of the quadratures of each of the individual oscillators, as well as the 4$\times$4 covariance matrix for all four quadratures. We define the $X$ and $Y$ quadrature operators as $X_a = (a+a^\dagger)/2$, $X_b = (b+b^\dagger)/2$, $Y_a = (a-a^\dagger)/(2i)$, and $Y_b = (b-b^\dagger)/(2i)$. Further, defining the vectors $\mathbf{v}_a^{\ms{T}} = (X _a,  Y _a)$ and $\mathbf{v}_b^{\ms{T}} = (X _b,  Y _b)$, the covariance matrices for each single oscillator are, respectively, 
\begin{align}
    C_a =  \langle \mathbf{v}_a \mathbf{v}_a ^{\ms{T}} \rangle - \langle \mathbf{v}_a \rangle \langle\mathbf{v}_a ^{\ms{T}} \rangle,
     \\
     C_b =  \langle \mathbf{v}_b \mathbf{v}_b ^{\ms{T}} \rangle - \langle \mathbf{v}_b \rangle \langle\mathbf{v}_b ^{\ms{T}} \rangle.
\end{align}
The eigenvectors of the covariance matrices give the quadratures (the combinations of $X$ and $Y$) that have maximum and minimum variances (and thus yield, respectively, the highest squeezing or the highest amplification), and the corresponding eigenvalues give these minimum and maximum variances. We denote the minimum and maximum variances for the respective modes by $V_a^{\ms{min}}$,  $V_a^{\ms{max}}$, $V_b^{\ms{min}}$, and $V_b^{\ms{max}}$. 

Similarly, the covariance matrix for the full two-mode system is given by 
\begin{align}
    C =  \langle \mathbf{v}\mathbf{v}^{\ms{T}} \rangle - \langle \mathbf{v}\rangle \langle\mathbf{v}^{\ms{T}} \rangle 
\end{align}
with $\mathbf{v}^{\ms{T}} \equiv (X_a, Y_a, X_b, Y_b)$. In general, the quadratures given by the eigenvectors of this covariance matrix are the linear combinations of all four of the $X$ and $Y$ quadratures and, therefore, do not belong to a single oscillator. A squeezed quadrature of this type, in general, describes quantum correlations between the two oscillators, and is referred to as two mode squeezing. As we elucidate below, the linear evolution of the coupled oscillators can convert two-mode or ``joint" squeezing into independent squeezing of both oscillators. We denote the maximum and minimum quadrature variances for the two-mode system by $V^{\ms{max}}$ and $V^{\ms{min}}$, respectively.  

From Eq.~(\ref{Lindb_eq}), we can obtain the equation of motion for the two-mode covariance matrix, which is 
\begin{equation}\label{eq14}
    \frac{\mathrm d C}{\mathrm d t} = A(t) C + C A^{\ms{T}}(t) + G,
\end{equation}
where 
\begin{equation}
    A(t) = 
     \begin{pmatrix}
        - \dfrac{\gamma}{2} & \omega_{\ms{c}} & 0 & 0 \\
        - \omega_{\ms{c}} & -\dfrac{\gamma}{2} & -2 g & 0 \\
        0  & 0  & - \dfrac{\kappa}{2} & \Omega(t)  \\
        -2g & 0 & -\Omega(t) & - \dfrac{\kappa}{2} 
    \end{pmatrix}, 
\end{equation}

\begin{equation}
    G = \frac{1}{4}
     \begin{pmatrix}
       \gamma(2 n_T + 1) & 0 & 0 & 0 \\
        0 & \gamma(2 n_T + 1) & 0 & 0 \\
        0  & 0  & \kappa(2 n_{\mathrm{s}} + 1) & 0  \\
        0 & 0 & 0 & \kappa(2 n_{\mathrm{s}} + 1)
    \end{pmatrix} , 
\end{equation}
and $\Omega(t) = \omega_{\ms{s}} + \Lambda \sin{\omega t}$. 

Since we may have to deal with large amounts of amplification and squeezing, a logarithmic scale is useful to quantify them. In this work, we quote decibels using the quadrature standard deviation, i.e., the square root of the quadrature variance. We set the zero of this scale to the coherent-state standard deviation, \(\sqrt{V_0}=1/2\). Hence, if the variance of a quadrature is \(V\), then
\begin{align}\label{eq:dBconv}
    \mathcal{S}\left(V\right)
    =
    10\log_{10}\left(\sqrt{\frac{V}{V_0}}\right)
    =
    10\log_{10}\left(2\sqrt{V}\right).
\end{align}
Similar logarithmic conventions based on the quadrature
standard deviation have been used in related quantum-metrology and
squeezing contexts~\cite{Reilly2024PDD,Vybornyi2026LineSearch}. The convention employing the quadrature variances  commonly used for squeezed-light noise powers,
\(10\log_{10}(V/V_0)\), differs from Eq.~(\ref{eq:dBconv}) by a factor of two.

An oscillator (or any number of coupled oscillators) can only be said to be squeezed if the minimum of the quadrature variances over all quadratures, $V_{\ms{min}}$, is less than $V_0$. We therefore define the squeezing of a set of oscillators as 
\begin{align} 
    \mathcal{S}_{\ms{sqz}} \equiv -\min \left[ 0, \mathcal{S}(V_{\ms{min}}) \right] , 
\end{align}
and, similarly, the amplification of a set of oscillators, assuming they began in coherent states, as  
\begin{align}
    \mathcal{S}_{\ms{amp}} \equiv \max \left[  0, \mathcal{S}(V_{\ms{max}}) \right] . 
\end{align}
In fact, since the maximum variance over all quadratures can never be less than $V_0$, we can simply set $\mathcal{S}_{\ms{amp}} = \mathcal{S}(V_{\ms{max}})$.

\section{Perturbative analysis} 
\label{pert}

In the regime in which the detuning between the spins and the cavity mode is much greater than the cavity/spin collective coupling rate, the NV centers shift the cavity frequency in what is often referred to as the ``dispersive'' interaction regime~\cite{Gard24}. Modulating the frequency of the NV centers therefore modulates the detuning and this, in turn, modulates the frequency of the cavity. Since it is known that modulating the cavity frequency at twice its frequency produces amplification and squeezing, we focus on the dispersive regime. 

To determine quantitatively the effect of parametric driving of the spins on our coupled system, we perform numerical simulations. Before we do so we undertake a perturbative analysis to provide insight into the parameter values that will give maximum amplification and squeezing. This analysis shows that one way to achieve this maximum is to choose the modulation frequency to be the sum of the cavity and the spin frequencies.  

We employ the time-dependent perturbation theory described in Ref.~\cite{Denys_ArXiv_2025}, which can be considered a version of time-independent perturbation theory for time-dependent Hamiltonians: a time-dependent Hamiltonian $H_0$ (for which the evolution operator $U^{(0)}(t,0)$ can be obtained) is perturbed by another, in general, time-dependent Hamiltonian $V$. The perturbation theory provides an expression for the full evolution operator, $U(t,0)$, as a series in which successive terms are proportional to increasing powers of $V$. This is appropriate for our system because we consider the regime in which the interaction rate $g$ is much smaller than the frequencies $\omega_{\ms{c}}$, $\omega_{\ms{s}}$, and the detuning $\Delta=\omega_{\ms{s}}-\omega_{\ms{c}}$. For the system under study, the (rate) perturbation Hamiltonian $\tilde{V} = g (a + a^\dagger)(b^\dagger + b)$ is time-independent, while $\tilde{H}_0$, being the rest of the rate Hamiltonian $\tilde{H}_{\mbox{\scriptsize HP}}$, is time-dependent. 

The evolution operator can be written as  
\begin{align}
    U(t) = U^{(0)}(t) + g U^{(1)}(t) + g^2 U^{(2)}(t) + \cdots . 
\end{align}
in which $U^{(0)}(t,0)$ is the solution to 
\begin{align}
   i \frac{d}{dt}  U^{(0)}(t) = \tilde{H}_0(t) U^{(0)}(t),
\end{align}
while the linear term is~\cite[Eq.~(3.40)]{Denys_ArXiv_2025} 
\begin{align}
    {U}^{(1)}(t) &= - i \int_{0}^{t} \!\! d\tau_1 \, {U}^{(0)}(t, \tau_1)  {\tilde{V}}(\tau_1) {U}^{(0)}(\tau_1, 0). 
\end{align} 

To express the evolution operators, we explicitly write their matrix elements for the spin subsystem, which are, therefore, operators in the cavity mode subsystem. In particular, we label the matrix elements of the spin $N/2$ system, which has $N+1$ states, with $s$ and $s'$, in which each takes the $N+1$ values $-N/2, -N/2+1, \ldots, N/2$. Using this notation, the matrix elements of $U^{(0)}(t)$ are 
\begin{equation}
     U^{(0)}_{ss'}(t) = \delta_{ss'}\exp\left\{-i \left[(\omega_c  a^\dagger  a + s\omega_{\ms{s}}) t - \frac{s \Lambda}{\omega}(\cos \omega t - 1)\right]\right\},
\end{equation}
where $\delta_{ss'}$ is a Kronecker delta. The first-order correction has the matrix elements:
\begin{equation}
     U^{(1)}_{ss'}(t) =  U^{(1)}_1(t; s) \, \delta_{s + 1, s'} +  U^{(1)}_2(t; s) \, \delta_{s, s' + 1},
\end{equation}
where
\begin{widetext}
\begin{align}
     U^{(1)}_1(t; s) = -\lambda C(s) 
     \left( \begin{array}{c}
        a e^{-i\omega_c t}  \\  a^\dagger e^{i\omega_c t}
     \end{array}\right)^{\ms{T}} 
     \left( \begin{array}{c}
        F(-\Sigma, -\Lambda, t)  \\  F( -\Delta, -\Lambda, t)
     \end{array}\right) 
    \exp\left\{i\left[ \left(\omega_{c}  a^\dagger  a  + \omega_{\ms{s}} s \right) t  -\frac{\Lambda}{\omega}(s \cos \omega t - s - 1)\right]\right\},
\end{align}
\begin{align}
     U^{(1)}_2(t; s) = -\lambda C(-s) 
          \left( \begin{array}{c}
        a e^{-i\omega_c t}  \\  a^\dagger e^{i\omega_c t}
     \end{array}\right)^{\ms{T}} 
     \left( \begin{array}{c}
          F( \Delta, \Lambda, t) \\ F(\Sigma, \Lambda, t)  
     \end{array}\right)
    \exp\left\{-i\left[ \left( \omega_{\ms{s}} s - \omega_{c}  a^\dagger  a \right) t -\frac{\Lambda}{\omega}(s \cos \omega t - s + 1) \right] \right\}, 
\end{align}
\end{widetext}
with $C(s) = \sqrt{(N/2+s+1)(N/2-s)}$ and 
\begin{align}\label{eq:res}
F(x,y, t) = \sum_{n = -\infty}^{\infty}
\begin{cases}
\displaystyle i^n J_n\left(-\dfrac{y}{\omega} \right) \frac{\left[ e^{i (x + n\omega)t} - 1 \right]}{x + n\omega}, & \quad x + n\omega \neq 0;
\\  &  \\
i^{n+1} J_n\left(-\dfrac{y}{\omega} \right) t, & \quad x + n\omega = 0.
\end{cases}
\end{align}
Here, as above, the spin/cavity detuning is denoted by $\Delta = \omega_{\ms{s}}- \omega_c$ and we have introduced $\Sigma = \omega_{\ms{s}} + \omega_c$ as the sum of the spin and cavity frequencies. 

Resonant phenomena are evident in the denominator of the function $F$. In particular, we have resonance conditions when the parametric drive frequency $\omega$ is an integer divisor of either the sum or difference of the spin and cavity frequencies:
\begin{align}
    \omega =  \frac{|\omega_{\ms{s}} \pm \omega_{\ms{c}}|}{n}, \;\;\; n \in \mathbb{N} . 
\end{align}
Therefore, we look for amplification and squeezing when $\omega$ is close to $\Sigma$ and/or $|\Delta|$.

\section{Performance of the parametric amplifier}
\label{nums}

The numerical results in this section are obtained from the linear covariance equation, Eq.~(\ref{eq14}). The dynamics has two very different timescales: the fast oscillations of the oscillators and the much slower amplification or squeezing rate. The equations are therefore ``stiff" and require very high accuracy on the short timescale to obtain sufficient accuracy on the long timescale. We achieve this by constructing the one-period covariance map by integrating Eq.~(\ref{eq14}) over a single period with adaptive step-size control. We then iterate this map to obtain the stroboscopic covariance dynamics over many modulation periods. The code used to generate the numerical results is available in Ref.~\cite{CodeRepository}.

\begin{figure}[t]
    \centering
\includegraphics[width=1\columnwidth]{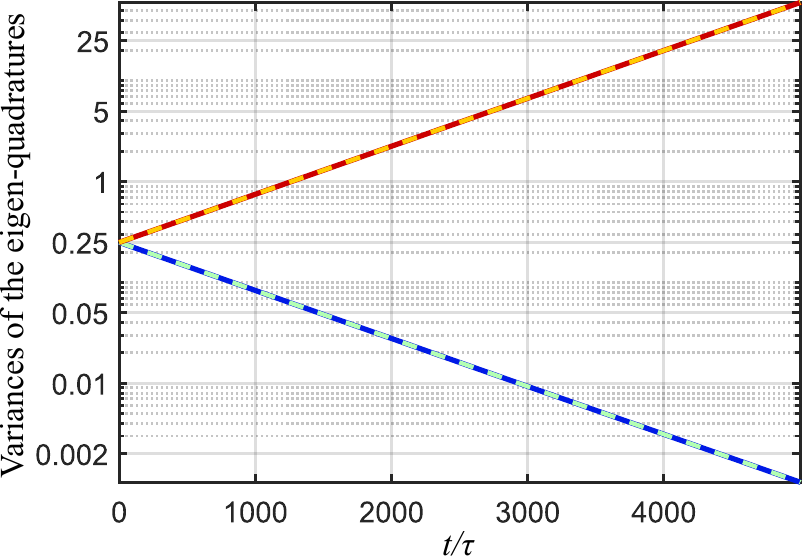} 

    \caption{Evolution of the eigenvalues of the two-mode covariance matrix under modulation of the frequency of the ensemble of nitrogen-vacancy centers coupled to a cavity mode. The modulation frequency~$\omega$ is the sum of the cavity mode frequency and the NV transition frequency. Here we assume ideal conditions in which neither the cavity nor the spins are damped. The covariance matrix has four eigenvalues corresponding to the variances of four quadratures. These eigenvalues are plotted respectively as: orange solid, red dashed, light blue solid, and dark blue dashed lines. Two quadratures are amplified at the same rate, while the other two are correspondingly squeezed. The period of the modulation is $\tau = 2\pi/\omega$.}
    \label{fig:Vs_ev}
\end{figure}

For our numerical analysis, unless specified otherwise, 
we take $\omega_{\ms{c}}/(2\pi) = 1.5~\mbox{GHz}$, $\omega_{\ms{s}}/(2\pi) = 0.5~\mbox{GHz}$, parametric driving frequency $\omega = \omega_{\ms{c}} + \omega_{\ms{s}} = 2\pi \,\times \, 2~\mbox{GHz}$, collective spin/cavity coupling rate $g/(2\pi) = 3.5~\mbox{MHz}$, and the cavity and effective spin damping rates, $\gamma/(2\pi) = \kappa/(2\pi) = 200~\mbox{kHz}$. The cavity output coupling and internal loss rates, respectively, are $\gamma_{\mathrm{c}}=\gamma_{\mathrm{l}}= 2\pi \times 100~\mbox{kHz}$. 

We first examine the evolution at the resonance $\omega = \omega_{\ms{c}} + \omega_{\ms{s}}$ under ideal conditions. That is, we simulate the evolution of the Hamiltonian~(\ref{Ham}) without damping of the cavity or the spin subsystem and at zero temperature. In Fig.~\ref{fig:Vs_ev}, we plot the evolution of the eigenvalues of the two-mode covariance matrix, which are the variances of the eigen-quadratures. The eigen-quadratures are the quadratures of the two-mode system that are uncorrelated with each other. They also have the property that two of them respectively possess the maximum and minimum variances over all possible linear combinations of the four quadratures, and, thus, reveal any amplification or squeezing generated by the evolution.  

We see from Fig.~\ref{fig:Vs_ev} that the rate of amplification and squeezing is essentially constant under the parametric drive. Note that there are two quadratures that are amplified by essentially the same amount, and thus two quadratures that are correspondingly squeezed by the same amount. These four eigen-quadratures are not quadratures of either of the two physical oscillators but of oscillators that are linear combinations of the two. The squeezing thus indicates entanglement between the two systems. However, we show in Sec.~\ref{1modeSqz} that by using two simple operations---evolution with suitably chosen detuning between the systems and evolution under the coupling between the spins and the cavity when on resonance (a partial Rabi oscillation)---the squeezing can be transformed in such a way that it becomes one-mode squeezing simultaneously for both the cavity and the spins. It can also be transformed so that it is squeezing in two fifty/fifty linear combinations of the oscillators meaning that the cavity and the spins are maximally entangled for that level of squeezing (pure two-mode squeezing).

\subsection{Amplification}

\begin{figure}[t]
    \centering
\includegraphics[width=1\columnwidth]{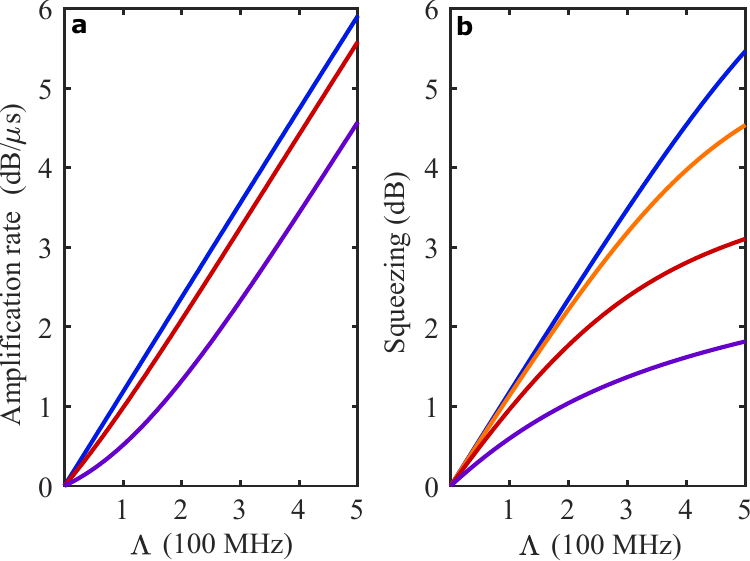} 
    \caption{(a) Amplification rate as a function of the NV modulation amplitude~$\Lambda$ when the spin modulation frequency~$\omega = \omega_{\ms{s}} + \omega_{\ms{c}}$. While this amplification rate is essentially independent of temperature it does depend on the damping rates of the cavities. Blue: $\gamma / (2\pi) = \kappa / (2\pi)  = 2~\mbox{kHz}$. Red: $\gamma / (2\pi) = \kappa / (2\pi) = 50~\mbox{kHz}$. Purple: $\gamma / (2\pi) = \kappa / (2\pi) = 200~\mbox{kHz}$. (b) Maximum squeezing of the joint quadratures of the cavity mode and NV ensemble as a function of the modulation amplitude $\Lambda$ at a temperature of $T=10~\mbox{mK}$. Blue: $\gamma / (2\pi) = \kappa / (2\pi) = 2~\mbox{kHz}$. Orange: $\gamma / (2\pi) = \kappa / (2\pi) = 10~\mbox{kHz}$. Red: $\gamma / (2\pi) = \kappa / (2\pi) = 50~\mbox{kHz}$. Purple: $\gamma / (2\pi) = \kappa / (2\pi) = 200~\mbox{kHz}$. }
    \label{fig:amp_and_sqz}
\end{figure}

\begin{figure}[t]
    \centering
    \includegraphics[width=1\columnwidth]{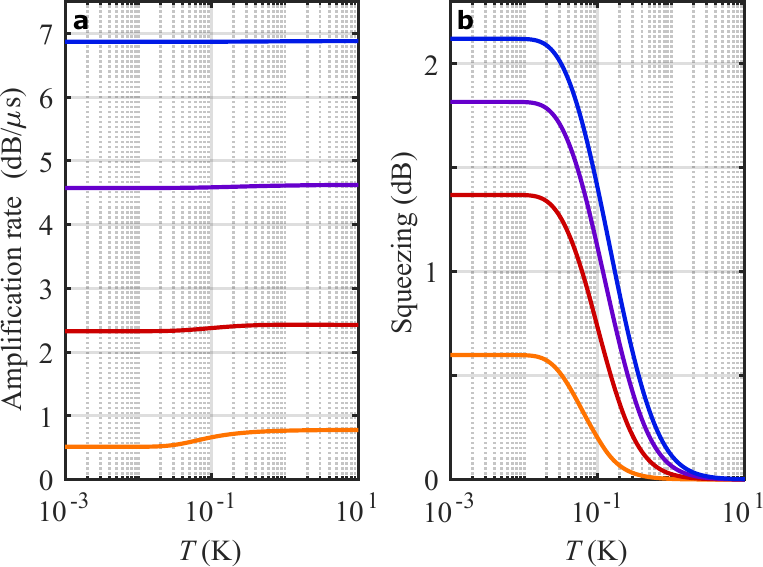} 
    \caption{ a) Amplification rate generated by parametric driving of the NV ensemble as a function of temperature for four values of the driving amplitude, $\Lambda$. There is essentially no change in the amplification rate between 10 K and 300 K. Blue: $\Lambda/(2\pi) = 700~\mbox{MHz}$. Purple: $\Lambda/(2\pi) = 500~\mbox{MHz}$. Red: $\Lambda/(2\pi) = 300~\mbox{MHz}$. Orange: $\Lambda/(2\pi) = 100~\mbox{MHz}$.  b) Maximum squeezing generated by parametric driving of the NV ensemble as a function of temperature for the same four values of the driving amplitude, $\Lambda$. No squeezing is generated above 1 K. As above, the parametric drive $\omega = \omega_{\ms{c}} + \omega_{\ms{s}}$, where $\omega_{\ms{c}}$ is the cavity mode frequency and $\omega_{\ms{s}}$ is the transition frequency of the NV spins. 
    }
    \label{fig:temp}
\end{figure}

From Fig.~\ref{fig:Vs_ev} we see that the driven unitary evolution of the system continually squeezes (and thus amplifies) two linear combinations of the modes at a constant rate. We now examine how this rate changes as a function of the parametric driving amplitude, as well as the effect of that cavity and spin decay and nonzero temperature. In Fig.~\ref{fig:amp_and_sqz}a, we plot the rate of amplification of the amplified eigen-quadratures, in decibels per microsecond, as a function of the amplitude of the parametric drive $\Lambda$. We note that both the cavity mode and the NV ensemble are amplified at essentially the same rate as that of the eigen-quadratures, so we do not plot these separately. Essentially, for the purposes of amplification, it does not matter that the squeezing is joint (two-mode) squeezing of the oscillators; both oscillators are amplified, although neither is individually squeezed. In Fig.~\ref{fig:amp_and_sqz}a, we show the rate of amplification at a temperature of $T=10~\mbox{mK}$, without any damping of the two modes and when they are both damped with $\gamma/(2 \pi) = \kappa /(2 \pi) = $ $50~\mbox{kHz}$ and $200~\mbox{kHz}$ (the red and purple curves, respectively). We find that increasing the temperature to $300~\mbox{K}$ does not change the amplification rates appreciably. In Fig.~\ref{fig:temp}a, we plot the amplification rate as a function of temperature from 1~mK to 10~K, and for four values of the driving amplitude. This similarly shows that the amplification is largely unaffected by temperature. 

We also examined the bandwidth of the amplification by fixing the modulation frequency at $2~\mbox{GHz}$, the amplification time at $2~\mu\mbox{s}$, and varying the cavity frequency. We find that the amplification reduces as the cavity frequency moves away from the resonance condition. We define the amplification bandwidth as the FWHM; that is, the separation between the two frequencies at which the growth rate falls to half of its maximum value. For the nominal parameters used here, this gives a bandwidth of approximately $500~\mbox{kHz}$, and within the parameter range tested it does not appear to be significantly affected by the damping rates, the temperature, or the driving amplitude.

The scale of the amplification bandwidth can be understood from the perturbative expression
in Eq.~(\ref{eq:res}). Near the sum-frequency resonance, the dominant term has
\(x=-\Sigma\) and \(n=1\), so that
\begin{equation}
    x+n\omega=\omega-(\omega_{\ms{c}}+\omega_{\ms{s}})
    \equiv \delta_{\ms{r}} .
\end{equation}
The resonant factor in Eq.~(\ref{eq:res}) is then
\begin{equation}
    \frac{e^{i\delta_{\ms{r}}t}-1}{\delta_{\ms{r}}}
    =
    i t e^{i\delta_{\ms{r}}t/2}
    {\rm sinc}\!\left(\frac{\delta_{\ms{r}}t}{2}\right).
\end{equation}
Thus the finite-time resonance response is Fourier limited. If we define FWHM in terms of the amplitude, then for \(t=2~\mbox{$\mu$s}\) FWHM given by the sinc functions is approximately \(0.6~\mbox{MHz}\). This is similar to our numerically extracted value and explains why the bandwidth is not primarily set by the damping rates in the parameter range considered here.

\subsection{Steady-state gain and noise temperature}

To obtain a good estimate of the noise temperature of the amplifier when operated in the steady-state, we first note that evolution of the amplification and squeezing generated by the parametric driving is exponential, which is the same as that generated by a non-degenerate parametric amplifier (see Sec.~\ref{tmsq} below). We thus replace the frequency modulation in the Hamiltonian $\tilde{H}_{\mathrm{HP}}$ with the Hamiltonian for the non-degenerate paramp and perform an analytic steady-state analysis of the gain and the noise temperature. We present this analysis in the appendix which gives the steady-state gain of the amplifier as
\begin{align}
    G_{\mathrm{ss}} =  \frac{ (\gamma_{\mathrm{c}} - \gamma_{\mathrm{l}}) \kappa + k^2 }{(\gamma_{\mathrm{c}} + \gamma_{\mathrm{l}})\kappa -  k^2 } = \frac{ (\gamma_{\mathrm{c}} - \gamma_{\mathrm{l}})/\gamma + (k/\!\sqrt{\gamma\kappa})^2 }{1 -  (k/\!\sqrt{\gamma\kappa})^2 } . 
\end{align}
The quantity $k$ is the rate constant of the fictitious non-degenerate paramp used to model the spin modulation and is related to the rate of amplification generated by the spin modulation in the absence of any damping, in dB's per second, $\dot{S}(V)$, by $k  \approx 0.46 \,\dot{S}(V)$ (see Appendix~\ref{app}).

To achieve high gain in the steady-state, one must set the paramp rate $k$ slightly below the geometric mean of the damping rates $\gamma$ and $\kappa$. As $k$ approaches $\!\sqrt{\kappa\gamma}$, the steady-state gain tends to infinity. The system becomes unstable once $k \geq \!\sqrt{\kappa\gamma}$. 
We find that with $\kappa = \gamma = 2\pi\times 200~\mbox{kHz}$ a modulation amplitude of $335~\mbox{MHz}$ puts $k$ close to $\!\sqrt{\kappa\gamma}$. 

If we set the input/output coupling to be much larger than the NV loss rate ($\gamma_{\mathrm{l}} \ll \gamma$) and the gain to be high ($k^2/(\gamma\kappa) \sim 1$), then an approximate expression for the gain is 
\begin{align}
    G_{\mathrm{ss}} \approx \frac{ 2 }{1 -  \xi^2 } . 
\end{align}
where we have defined $\xi \equiv k/\!\sqrt{\gamma\kappa}$.

The added noise power per unit frequency at the input, and at the cavity frequency in units of photons, can be written as  
\begin{align}
    \mathcal{N}_{\mathrm{add}} = n_{\mathrm{add}} + \frac{1}{2},
\end{align}
where
\begin{align}
    n_{\mathrm{add}} & = \eta \frac{ (\gamma_{\mathrm{1}}/\gamma) (2 n_{T} +1)   + \xi^2  (2 n_{\mathrm{s}}  +1)  }
    { [\eta  +
    \xi^2 - \gamma_{\mathrm{l}}/\gamma ]^2 }  -\frac{1}{2} . \label{nadd}
\end{align}
is the number of photons of noise added above the vacuum level of $1/2$, and  \begin{align}
    \eta \equiv \gamma_{\mathrm{c}}/\gamma. 
\end{align}
As above, $n_T$ is the number of thermal microwave photons at the ambient temperature of the cavity and $n_{\mathrm{s}}$ is the number of thermal photons at the temperature of the spins. At 80\% polarization the spin temperature is $\sim 11~\mbox{mK}$ resulting in $n_{ \mathrm{s}} \sim 1/8$. 

The expression for $n_{\mathrm{add}}$ in Eq.~(\ref{nadd}) may look rather complex at first glance. But if we once again take the steady-state gain to be much larger than unity ($k \sim \sqrt{\kappa\gamma}$ or $\xi \approx 1$) and we impedance match the cavity ($\gamma_{\ms{c}} = \gamma_{\ms{l}}$), we have $\eta = 1/2$ and the number of added noise photons at the input reduces to 
\begin{align}
    n_{\mathrm{add}} & \approx \frac{1}{2} n_{T}   +  n_{\mathrm{s}}  + \frac{1}{4}, \label{naddsimple}
\end{align}
which can be much lower than that given by the ambient temperature, $T$. In terms of $n_{\mathrm{add}}$ the noise temperature of the amplifier is 
\begin{align}
     T_{\mathrm{amp}}  = \frac{\hbar\omega_{\mathrm{c}}}{k_{\mathrm{B}}} \left(\ln \left[ \frac{1}{n_{\mathrm{add}}} + 1 \right]\right)^{-1} . 
\end{align}
Within the parametric amplifier model of our system the spins \textcolor{red}{and} the cavity mode have the same frequency. Thus in terms of the spin polarization the temperature of the spins is 
\begin{align}
       T  & = \frac{\hbar\omega_{\ms{c}}}{k} \left( \ln \left[ \frac{1 + P}{1 - P} \right] \right)^{-1} 
\end{align}
and the resulting value of $n_{\ms{s}}$ in terms of the spin polarization is 
\begin{align}
    n_{\ms{s}} = \frac{1}{2}\frac{1 - P}{P}.
\end{align}

At room temperature (300 K) with $\omega_{\ms{c}}/(2\pi) = 1.5~\mbox{GHz}$ and the spins pumped to 80\% polarization we have $n_T = 4160$ and $n_{\ms{s}} = 1/8$. From Eq.~(\ref{naddsimple}) the resulting added noise in photons at the amplifier input is 
\begin{align}
    n_{\mathrm{add}} & \approx \frac{1}{2} n_{T}   +  n_{\mathrm{s}}  + \frac{1}{4}  \approx  \frac{1}{2} n_{T} = 2080  
\end{align} 
and the noise temperature is 
\begin{align} 
    T_{\mathrm{amp}}  = 150~\mbox{K} .
\end{align}
The noise temperature is essentially the same as that of the maser amplifier demonstrated in~\cite{day2024maseramp}. Note that the noise temperature of the amplifier can be significantly lower than the ambient temperature (e.g., room temperature) because of the cooling provided by the spins. The noise temperature of the amplifier is set by the ambient temperature and the parameters that determine the rate of cooling. Thus the colder the ambient temperature the lower the noise temperature.  

\subsection{Two-mode squeezing}
\label{tmsq}
The squeezing of the eigen-quadratures generated by the parametric drive at $10~\mbox{mK}$ reaches a maximum within 3 $\mu s$ and then decreases somewhat to a steady-state. We plot the maximum value in Fig.~\ref{fig:amp_and_sqz}b as a function of the driving amplitude~$\Lambda$, and for four values of the cavity and spin damping rates. We see that to achieve large squeezing, we need very high-$Q$ cavities. At this time, a realistic value for the cavity damping rate is $200~\mbox{kHz}$, which gives about 1.8 dB of squeezing at a modulation amplitude of $\Lambda/(2\pi)=500~\mbox{MHz}$ (the purple curve in Fig.~\ref{fig:amp_and_sqz}b).  

In Fig.~\ref{fig:temp}b, we plot the maximum squeezing as a function of temperature with cavity and spin damping rates at $200~\mbox{kHz}$. We see that essentially no squeezing is achieved above 1 K.  

It is also important to examine the quadratures in which the squeezing appears. For the 1.8 dB squeezing achieved with a modulation amplitude of $\Lambda/(2\pi)= 500~\mbox{MHz}$ at $T = 10~\mbox{mK}$, the two squeezed quadratures are 
\begin{align}
    Q_1 & \approx 0.7 X_a + 0.03 X_b + 0.7 Y_a + 0.13 Y_b , \label{Q1} \\
    Q_2 & \approx 0.03 X_a + 0.71 X_b + 0.13 Y_a - 0.69 Y_b , \label{Q2}
\end{align}
neither of which are quadratures of a single oscillator. Thus, neither oscillator is squeezed in isolation; we have a more general form of two-mode squeezing. We note that \textit{maximal} two-mode squeezing refers to a situation in which, for a given value of squeezing, at least one quadrature of one oscillator is maximally correlated (in this case, also entangled) with a quadrature of the other oscillator. This means that one of the maximally squeezed quadratures can be written as  
\begin{align}
    Q & = \frac{1}{\sqrt{2}} ( x_a X_a + y_a Y_a ) + \frac{1}{\sqrt{2}} ( x_b X_b + y_b Y_b ),
\end{align}
where $x_a^2 + y_a^2 = x_b^2 + y_b^2 = 1$. ($Q$ contains an equal weighting of a quadrature of each oscillator.) 
In the next section, in addition to considering how to prepare single-mode squeezing of both oscillators, we show that the standard operations available in this two-mode system can be used to adjust the squeezing so that it achieves maximal two-mode squeezing (i.e., maximal two-mode entanglement for the given level of squeezing).

\begin{figure}[t]
    \centering
\includegraphics[width=1\columnwidth]{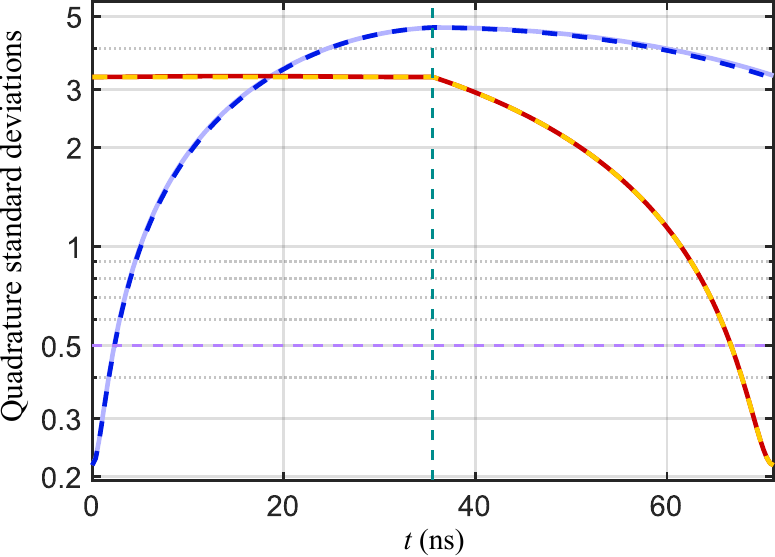} 
    \caption{The evolution of the squeezing of four quadratures under a simple two-part control protocol that transfers two-mode (joint) squeezing generated by the modulation of the NVs into simultaneous independent single-mode squeezing of the cavity mode and the NV ensemble. The standard deviations of the two quadratures that are initially squeezed are plotted in light blue (solid) and dark blue (dashed). The single-mode quadratures to which the squeezing is transferred are plotted in orange (solid) and red (dashed). The horizontal purple dashed line is the standard deviation of a quadrature for a coherent state (the value for zero squeezing) and the vertical green dashed line gives the point where the protocol switches from evolution under detuning to evolution under the RWA interaction.}
    \label{fig:swap}
\end{figure}

\subsection{Single-mode squeezing of microwaves and spins}
\label{1modeSqz}

When we turn off the parametric driving, we can apply the rotating-wave approximation (RWA) and move into the interaction picture with respect to the free cavity Hamiltonian. The rate Hamiltonian becomes 
\begin{align}
    \tilde{H}_{\mathrm{RWA}} = \tilde{H}_{b}^{\mathrm{RWA}} + \tilde{H}_{\mathrm{int}}^{\mathrm{RWA}} 
\end{align} 
with 
\begin{align}
   \tilde{H}_{b}^{\mathrm{RWA}} & =     \Delta b^\dagger b ,
   \\
   \tilde{H}_{\mathrm{int}}^{\mathrm{RWA}} & = g (a b^\dagger + b a^\dagger) .
\end{align}
The Hamiltonian $\tilde{H}_{b}^{\mathrm{RWA}}$ generates a rotation in the space spanned by $X_b$ and $Y_b$, namely 
\begin{align}
   \left( \begin{array}{c}
     X_b(t)   \\ Y_b(t) 
   \end{array} \right) & = \left( \begin{array}{cc}
     \cos\phi(t)   & - \sin\phi(t) \\
      \sin\phi(t)  & \cos\phi(t)
   \end{array}\right)      
   \left( \begin{array}{c}
     X_b(0)  \\ Y_b(0) 
   \end{array} \right)  ,  
\end{align}
with $\phi = \Delta t$. The Hamiltonian $\tilde{H}_{\mathrm{int}}^{\mathrm{RWA}}$ generates a rotation in the space spanned by $X_a$ and $X_b$ and a simultaneous and identical rotation in the space spanned by $Y_a$ and $Y_b$:
\begin{align}
   \left( \begin{array}{c}
     X_a(t)   \\ X_b(t) 
   \end{array} \right) & = \left( \begin{array}{cc}
     \cos\theta(t)   & - \sin\theta(t) \\
      \sin\theta(t)  & \cos\theta(t)
   \end{array}\right)      
   \left( \begin{array}{c}
     X_a(0)  \\ X_b(0) 
   \end{array} \right) ,  \\
   \left( \begin{array}{c}
     Y_a(t)   \\ Y_b(t) 
   \end{array} \right) & = \left( \begin{array}{cc}
     \cos\theta(t)   & - \sin\theta(t) \\
      \sin\theta(t)  & \cos\theta(t)
   \end{array}\right)      
   \left( \begin{array}{c}
     Y_a(0)  \\ Y_b(0) 
   \end{array} \right) , 
\end{align}
with $\theta = g t$.

We can write an arbitrary two-mode quadrature as 
\begin{align}
    Q = x_a X_a + y_a Y_a + x_b X_b + y_b Y_b,
\end{align}
where $x_a$, $x_b$, $y_a$, and $y_b$ are real coefficients. 
To transform arbitrary two-mode squeezing into single-mode squeezing of the two oscillators, we first note that if 
\begin{align}
    \frac{x_a}{y_a} = \frac{x_b}{y_b},
\end{align}
then we can write the quadrature as 
\begin{align}
    Q = \cos\!\theta\, (\cos\!\phi\, X_a - \sin\!\phi\,  Y_a) - \sin\!\theta\,(\cos\!\phi\, X_b - \sin\!\phi\, Y_b) .
    \label{Qform}
\end{align}
   Because the RWA interaction Hamiltonian transforms the $X$'s in the same way as it does the $Y$'s, when $Q$ is in the above form the application of $\tilde{H}_{\mathrm{int}}^{\mathrm{RWA}}$ for a time $t$ simply changes the value of $\theta$ to $\theta + gt$. One can therefore use the interaction between the oscillators to make $\theta$ zero (or $\pi/2$) thus placing all of the squeezed quadrature $Q$ into a single oscillator. When one maximally squeezed quadrature lies in one oscillator, the other squeezed quadrature (if there is one) necessarily lies in the other oscillator, and so both oscillators are simultaneously squeezed. 

In general, any two-mode quadrature $Q'$ can be written as 
\begin{align}
    Q' = \cos\!\theta\, (\cos\!\phi\, X_a - \sin\!\phi\,  Y_a) - \sin\!\theta\,(\cos\!\psi\, X_b - \sin\!\psi\, Y_b) . 
\end{align}
The application of $\tilde{H}_{b}^{\mathrm{RWA}}$ for a time $t$ merely changes $\psi$ to $\psi + \Delta t$. We can thus use $\tilde{H}_{b}^{\mathrm{RWA}}$ to rotate the spin oscillator so that $\psi = \phi$, placing the squeezed quadrature in the form given by Eq.~(\ref{Qform}). We can therefore transform any general two-mode squeezing into the squeezing of at least one of the oscillators by first evolving the system under a specified detuning $\Delta$ for a specified time, and then evolving it under the interaction between the two oscillators for another specified time. 

In Fig.~\ref{fig:swap} we show the evolution of the standard deviations of the single-mode quadratures under the two-part control described above. Starting with the two-mode squeezed quadratures $Q_1$ and $Q_2$ given in Eqs.~(\ref{Q1}) and (\ref{Q2}), the evolution under the detuning rotates the spin oscillator through an angle $\Delta\psi = 0.95\pi$, and the interaction performs a partial Rabi oscillation through an angle $\Delta\theta = 0.25\pi$. Given $g/(2\pi) = 3.5~\mbox{MHz}$, the latter takes 36 ns, and we choose the detuning so that the first step takes the same time. In Fig.~\ref{fig:swap} the standard deviations of the initially squeezed quadratures are plotted in dark and light blue, and the single-mode quadratures that become squeezed are plotted in orange and red. 

Finally, if we want instead to adjust the squeezing generated by the parametric driving so that it becomes maximal two-mode squeezing, then we first use $\tilde{H}_{b}^{\mathrm{RWA}}$ to transform one of the maximally squeezed quadratures so that it is in the form given by Eq.~(\ref{Qform}). Then we apply the interaction $\tilde{H}_{\mathrm{int}}^{\mathrm{RWA}}$ to change $\theta$ to $\pi/4$, so that the maximally squeezed quadrature contains an equal weighting of a quadrature of both oscillators. 

\section{Considerations for experimental realization}

\label{expreal}

We have elucidated the parameter regime required to realize a parametric amplifier by modulating the frequency of an ensemble of two-level emitters. We now consider what would be needed to achieve this parameter regime in future experiments. The requirements for realizing our parametric amplifier divide naturally into two parts: (i) achieving a cavity mode with the required high $Q$ and strong coupling to the emitters and (ii) achieving the modulation of the emitters without dissipating excessive power. Modulating the emitters is achieved by irradiating them with a sufficiently intense microwave field. To maximize the intensity of this field and minimize power dissipation, one must use a microwave cavity~\cite{Abhyankar20}. While the mode of this ``drive" cavity need not have as high $Q$ as that of the signal cavity, realizing the paramp nevertheless requires two cavity modes. One must either design a single cavity with two sufficiently high-$Q$ modes or design two overlapping cavities. Such configurations have not been explored extensively to-date, and thus constitute the primary unknown factor (and thus challenge) in realizing the paramp. While designing microwave cavity configurations is beyond our scope here, for reference we review what has been achieved to-date.

First, we consider the signal cavity mode which is required to achieve a high coupling constant $g$ and a high $Q$. The coupling constant scales as the square root of the product of the density of NV centers with the fraction of the mode volume that is filled with the NV diamond (the "filling fraction"). Current experiments typically use 3D dielectric resonators with a mode volume on the order of 1~cm$^3$, while NV diamonds are typically 1 mm thick and a few mm's across. This gives a rather low filling fraction resulting in $g/(2\pi)$ values of $200-300$~kHz~\cite{Wang24}. NV densities are limited to about $10^{17}$ ~cm$^{-3}$ to keep inhomogeneous linewidths of the order of a hundred kHz. This density at a filling fraction of unity would achieve a $g/(2\pi)$ of the order of 10 MHz. The size of the diamond tends to be limited by the attenuation of the pump laser; if the diamond is too thick, the laser is not able to access all the NV centers. To achieve larger filling fractions, one could either increase the total diamond volume and employ a creative geometry that minimizes laser attenuation, or reduce the mode volume of the cavity. Planar anapole resonators are potential candidates for taking the second route~\cite{Abhyankar20}. While arrays of anapole resonators have achieved $Q\sim10^5$, it is not yet clear whether this can be achieved within the constraints of an input-output coupled system integrated with NV diamond. 

We now turn to the question of achieving a reasonable power efficiency for the microwave drive that modulates the NV frequency. For EPR spectroscopy at room temperature, state-of-the-art 3D cavities provide a ``power conversion factor" of $C_{\ms{p}} = 0.75~\mbox{mT}/\!\sqrt{\mbox{W}}$~\cite{Haakon23, Abhyankar22}. Since the electron gyromagnetic ratio results in a frequency shift of  $28~\mbox{MHz}/\mbox{mT}$, this gives a modulation amplitude of $21~\mbox{MHz}/\!\sqrt{\mbox{W}}$. Exploring driving amplitudes at the low end, we find that with $\omega_{\ms{c}}/(2\pi) = 1.5~\mbox{GHz}$, $\omega_{\ms{s}}/(2\pi) = 0.5~\mbox{GHz}$ [thus $\omega/(2\pi) = 2~\mbox{GHz}$ and $\Delta /(2\pi) =-1~\mbox{GHz}$], $g/(2\pi) = 3.5~\mbox{MHz}$, $\gamma/(2\pi) = \kappa/(2\pi) = 200~\mbox{kHz}$, and a driving amplitude of $\Lambda/(2\pi) = 50~\mbox{MHz}$, one achieves an amplification rate of $220~\mbox{dB}/\mbox{ms}$. With the cavity power conversion factor given above, this would require about 5.7~W of microwave power. 

We note that much higher power conversion factors than those for 3D cavities have been achieved for planar anapole resonators, also at room temperature. Recent cavity designs have achieved $C_{\ms{p}} = 3~\mbox{mT}/\!\sqrt{\mbox{W}}$ and simulations show that $C_{\ms{p}} = 9~\mbox{mT}/\!\sqrt{\mbox{W}}$ should be rather feasible~\cite{Abhyankar20}. Using the latter value provides $252\!~\mbox{MHz}/\!\sqrt{\mbox{\small W}}$, so that an amplitude of $\Lambda/(2\pi) = 500~\mbox{MHz}$ would require slightly under 4~W of microwave power.  

The challenges to realizing parametric amplification, and thus squeezing, at cryogenic temperatures are somewhat different to those at room temperature. Much higher resonator quality factors and coupling rates $g$ can be achieved with present technology at cryogenic temperatures. However, a primary unknown factor in operating the paramp at cryogenic temperatures is the degree of quasiparticle excitation and heating that results from the microwave drive field and the laser that cools the NVs. Both these effects reduce resonator quality factors~\cite{Alexander25}. In view of this, it may well be necessary to set $Q$ of the drive cavity significantly lower than that of the signal cavity, so that the microwave drive dissipates primarily via the drive cavity.

In the absence of quasiparticle excitations or heating, for 3D aluminum microwave resonators quality factors achievable at cryogenic temperatures are well in excess of $10^7$~\cite{Kudra20}. While the introduction of an NV diamond crystal will reduce the quality factor, it can still be expected to be significantly higher than the quality factors of 3D resonators at room temperature. Superconducting stripline (planar) resonators do not achieve such high quality factors, but they do achieve very high coupling rates and very high power conversion factors. The stripline resonator in~\cite{Probst17} has an intrinsic damping rate of $200~\mbox{kHz}$ and achieves a very high power conversion factor of $C_{\ms{p}} = 30~\mbox{T}/\!\sqrt{\mbox{W}}$. This is due to the very high confinement of the field in the resonator. Further, since stripline resonators are planar it should be possible to achieve a high filling factor by placing a diamond slab or film on top of the resonator. 

\section{Conclusion}
\label{conc}

We have analyzed a theoretical proposal for an effective hybrid non-degenerate parametric amplifier based on an NV-center ensemble coupled to a microwave cavity mode. In contrast to the case of uncoupled NV centers, the effective hybrid amplification becomes feasible for an ensemble of NVs coupled dispersively to a single microwave cavity mode. Our numerical simulations suggest that in this case, parametric driving (modulation) of the NV frequency must be applied at the sum of the NV and cavity frequencies.

The non-degenerate paramp in the given driving regime amplifies and squeezes both systems. Due to the interaction between the systems, time-dependent control of the detuning between them can be used to transform the two-mode squeezing (entanglement) between the systems into individual squeezing of each system, providing a possible route to squeezing the collective state of an NV ensemble.

Given the above results, parametric driving of an ensemble of NVs or other microwave-frequency two-level systems has a number of potential applications. First, it can be viewed as a room-temperature microwave amplifier whose noise temperature can be far below the ambient temperature and comparable to that of optically pumped solid-state maser amplifiers, or equivalently as an amplifier for any signal stored in the NV ensemble. Second, so long as any heating and quasi-particle production from the microwave drive can be made sufficiently small, at low temperatures it could serve as a means of generating entanglement between an ensemble and a microwave field. In this same setting it might also provide a means to generate collective squeezing of an ensemble of spins, where the cavity mode serves as the conduit. 


We have also discussed implementation requirements and
feasibility estimates for a possible NV-cavity realization of the proposed
hybrid parametric amplifier. These include the quality factors, collective
spin-cavity coupling rates, and pump-field conversion factors needed in
cryogenic and room-temperature regimes. The primary implementation challenge is
the design of a single or dual microwave resonator that provides both a
relatively high-\(Q\) signal/idler mode and an efficient pump mode having a sizable spatial overlap with the addressed spin ensemble.

\textit{Acknowledgments.}
D.I.B. was supported by Army Research Office (ARO) (grant W911NF-23-1-0288; program manager Dr.~James Joseph). R.O. and A.G.S. acknowledge support by the National Research Foundation of Ukraine, project No.~2023.03/0073. The views and conclusions contained in this document are those of the authors and should not be interpreted as representing the official policies, either expressed or implied, of ARO, NSF, or the U.S. Government. The U.S. Government is authorized to reproduce and distribute reprints for Government purposes notwithstanding any copyright notation herein. 

\section*{Author contributions}

R.O. performed the simulations and the majority of the derivations, interpreted results, and wrote the majority of the first draft of the manuscript. D.I.B. assisted with simulations. K.J. performed some of the derivations and wrote part of the initial draft. M.E.T. conceived the project. D.I.B., K.J., and A.G.S. supervised the project and all authors analyzed the findings. D.I.B. and A.G.S. secured funding.  All authors discussed the results and refined the manuscript.

\section*{Competing interests}

The authors declare no competing interests

\section*{Additional information}

Correspondence and requests for materials should be addressed to Roman Ovsiannikov.

\appendix 
\section{Noise temperature of the amplifier}
\label{app}

Figure~\ref{fig:Vs_ev} shows that the parametrically driven Hamiltonian, Eq.~(\ref{Ham}), when driven at the sum frequency $\omega = \omega_{\mathrm{c}}+\omega_{\mathrm{s}}$ generates evolution that is very similar to that of a non-degenerate parametric amplifier. Hence, to estimate the noise temperature of the amplification, we can replace the parametric drive with a non-degenerate parametric amplifier and solve for the steady-state of the resulting linear two-mode system. First, the rate Hamiltonian $\tilde{H}_{\rm HP}$ is replaced by  
\begin{align}
    \tilde{H}_{\mathrm{ngp}} = -i \frac{k}{2} (a b - a^\dagger b^\dagger),
\end{align}
where we set $\Delta=0$ because the paramp operates with the two modes on resonance, and $k$ is proportional to the rate at which the original parametric driving of the NVs in $\tilde{H}_{\rm HP}$ amplifies the modes (we determine the constant of proportionality below in Eq.~(\ref{Randk})). If we consider the motion generated by $\tilde{H}_{\mathrm{ngp}}$ in the absence of damping, we have $\dot{a} =  (k/2) b^\dagger$ and $\dot{b}^\dagger = (k/2) a$. In terms of the $X$ quadratures this is $\dot{X}_a = (k/2) X_b$ and $\dot{X}_b = (k/2) X_a$. Defining $X_+ = X_a + X_b$ and $X_- = X_a - X_b$ we can rewrite these as 
\begin{align}
    \frac{d}{dt} X_+ & = (k/2) X_+  \rightarrow  X_+(t) = e^{k t/2} X_+, \nonumber \\
    \frac{d}{dt} X_- & = - (k/2) X_-  \rightarrow  X_-(t) = e^{-k t/2} X_-. \nonumber
\end{align}
The non-degenerate paramp thus creates maximal two-mode squeezing between the oscillators, which is thus very similar to that generated by our driven Hamiltonian $H_{\mathrm{HP}}$. The rate of amplification in decibels is 
\begin{align}
   R_{\mathrm{dB}} & = \frac{d}{dt} \mathcal{S}(V) = \frac{d}{dt} \left[ 10 \log_{10} \left(e^{k t/2} \right) \right] = \frac{d}{dt} \left[ 5 k t \log_{10}\left(e\right) \right] \nonumber \\
   & = 5 k  \log_{10}\left(e\right) \approx 2.2 k ,
   \label{Randk}
\end{align}
or $k \approx 0.46 \,\dot{S}(V)$. 

The quantum Langevin equations of input-output theory for the two oscillators, in terms of the $X$ quadratures, are~\cite{gardiner_input_1985,Jacobs14}
\begin{align}
    \dfrac{d}{dt} X_a & = -(\gamma/2) X_a + (k/2) X_b + \sqrt{\gamma_{\mathrm{c}}} A_{\mathrm{in}} + \sqrt{\gamma_{\mathrm{l}}} C_{\mathrm{in}}, \\
    \dfrac{d}{dt} X_b & = -(\kappa/2) X_b 
    + (k/2) X_a + \sqrt{\kappa} B_{\mathrm{in}}, 
\end{align}
where $\gamma = \gamma_{\mathrm{c}} + \gamma_{\mathrm{l}}$ with $\gamma_{\mathrm{c}}$ the coupling rate to the input field that is to be amplified and $\gamma_{\mathrm{l}}$ the internal loss rate of the cavity. The input fields are 
\begin{align}
    A_{\mathrm{in}}  = \frac{a_{\mathrm{in}} + a_{\mathrm{in}}^\dagger}{2} , \;\;
    B_{\mathrm{in}}  = \frac{b_{\mathrm{in}} + b_{\mathrm{in}}^\dagger}{2} , \;\; 
    C_{\mathrm{in}}  = \frac{c_{\mathrm{in}} + c_{\mathrm{in}}^\dagger}{2} ,
\end{align}
where $a_{\mathrm{in}}$ is the input field to be amplified, and $b_{\mathrm{in}}$ and $c_{\mathrm{in}}$ are the quantum noise sources associated with the spin and cavity losses, respectively. The correlation functions for these white noise sources are~\cite{gardiner_input_1985, Jacobs14}
\begin{align}
    \langle a_{\mathrm{in}}(t)a_{\mathrm{in}}^\dagger(t+t') \rangle & = (n_{\mathrm{T}} + 1) \delta(t') ,\\ 
    \langle b_{\mathrm{in}}(t)b_{\mathrm{in}}^\dagger(t+t') \rangle & =  (n_{\mathrm{s}}+1)\delta(t') ,\\
    \langle c_{\mathrm{in}}(t)c_{\mathrm{in}}^\dagger(t+t') \rangle & = (n_{\mathrm{T}} + 1) \delta(t') . 
\end{align}
Transforming to Fourier space the equations of motion become 
\begin{align}
   -i\nu  \left( \begin{array}{c}
       X_+  \\  X_-
    \end{array} \right) & = - \frac{1}{4}
     \left( \begin{array}{cc}
         \gamma + \kappa - 2k   & \gamma - \kappa \\ 
          \gamma - \kappa  &  \gamma + \kappa + 2k
    \end{array} \right)
     \left( \begin{array}{c}
       X_+  \\  X_-
    \end{array} \right)
    + \mathbf{X}_{\mathrm{in}}
\end{align}
with 
\begin{align}
   \mathbf{X}_{\mathrm{in}} =  
   \left( \begin{array}{c}
    \sqrt{\gamma_{\mathrm{c}}} A_{\mathrm{in}} + \sqrt{\gamma_{\mathrm{l}}} C_{\mathrm{in}} + \sqrt{\kappa} B_{\mathrm{in}}\\  
    \sqrt{\gamma_{\mathrm{c}}} A_{\mathrm{in}} + \sqrt{\gamma_{\mathrm{l}}} C_{\mathrm{in}} - \sqrt{\kappa} B_{\mathrm{in}}
    \end{array} \right),
\end{align}
where $\nu$ is the Fourier-space variable. The solution is 
\begin{align}
     \left( \begin{array}{c}
       X_+  \\  X_-
    \end{array} \right)
    & = \frac{\left( \begin{array}{cc}
         \gamma + \kappa -4i\nu + 2k  & \kappa - \gamma \\ 
          \kappa - \gamma  &  \gamma + \kappa - 4i\nu - 2k
    \end{array} \right) }
    {(\gamma + \kappa -4i\nu)^2/4 - (\kappa - \gamma)^2/4 - k^2  }\mathbf{X}_{\mathrm{in}}. 
\end{align}
The noise temperature of the amplifier is the noise added by the amplifier (that is, it does not include the noise that is part of the input signal being amplified). Further, while one must measure (or calculate) the total noise at the output of amplifier, the noise temperature is defined as the effective amount of noise that is added to the signal at the input. Thus one obtains the noise temperature by calculating the added noise at the output an dividing it by amplifier gain. We must first, therefore, determine the gain  (amplification) of the amplifier in the steady-state. To do so we compare the signal at the input, $A_{\mathrm{in}}$, to the signal at the output, $A_{\mathrm{out}} = \sqrt{\gamma_{\mathrm{c}}} X_a - A_{\mathrm{in}}$ to the input $A_{\mathrm{in}}$. At the cavity resonance ($\nu = 0$), and setting $B_{\mathrm{in}}$ and $C_{\mathrm{in}}$ to zero in the expressions for $X_+$ and $X_-$, we have 
\begin{align}
    A_{\mathrm{out}} = \frac{\sqrt{\gamma_{\mathrm{c}}}(X_+ + X_-)}{2} - A_{\mathrm{in}} =  \left( \frac{ (\gamma_{\mathrm{c}} - \gamma_{\mathrm{l}}) \kappa +  k^2 }
    {\gamma\kappa - k^2 } \right) A_{\mathrm{in}}.
\end{align} 
So that the steady-state amplifier gain is 
\begin{align}
    G_{\mathrm{ss}} =  \frac{ (\gamma_{\mathrm{c}} - \gamma_{\mathrm{l}}) \kappa + k^2 }{(\gamma_{\mathrm{c}} + \gamma_{\mathrm{l}})\kappa -  k^2 } . 
\end{align}
We see from this expression for the gain that when the driving rate $k$ exceeds $\sqrt{\gamma\kappa}$ the amplification overwhelms the damping and there is no steady-state. 

Now we need to determine the total noise power in the output $A_{\mathrm{out}}$. The output power spectrum is $S(\nu)$ where 
\begin{align}
   \frac{ \langle A_{\mathrm{out}}(\nu)A_{\mathrm{out}}(\nu') \rangle  +  \langle A_{\mathrm{out}}(\nu')A_{\mathrm{out}}(\nu) \rangle}{2}= S(\nu)\delta(\nu+\nu').
\end{align} 
We have 
\begin{align}
    A_{\mathrm{out}} & = \frac{   \left( 2\kappa -4i\nu \right)\sqrt{\gamma_{\mathrm{c}}\gamma_{\mathrm{l}}} C_{\mathrm{in}}   + k \sqrt{\gamma_{\mathrm{c}}\kappa} B_{\mathrm{in}}  }
    {\gamma\kappa - (4\nu^2+k^2) - 2i (\gamma + \kappa)\nu }  \\
    & \;\;\; + \frac{[ \kappa(\gamma_{\mathrm{c}}-\gamma_{\mathrm{l}}) + k^2 + 4 \nu^2 + 2i \nu (  \kappa + \gamma_{\mathrm{l}} -\gamma_{\mathrm{c}})   ] A_{\mathrm{in}}}{\gamma\kappa - (4\nu^2+k^2) - 2i (\gamma + \kappa)\nu} . \nonumber
\end{align}
For the purposes of calculating the noise added by the amplifier we need to discard any noise coming from the input, so we drop the term containing $A_{\mathrm{in}}$ to obtain 
\begin{align}
    A_{\mathrm{out}}^{\mathrm{add}} & = \frac{   \left( 2\kappa -4i\nu \right)\sqrt{\gamma_{\mathrm{c}}\gamma_{\mathrm{l}}} C_{\mathrm{in}}   + k \sqrt{\gamma_{\mathrm{c}}\kappa} B_{\mathrm{in}}  }
    {\gamma\kappa - (4\nu^2+k^2) - 2i (\gamma + \kappa)\nu } .
\end{align}
This gives us 
\begin{align}
    S(\nu) & = \left( \frac{\gamma_{\mathrm{c}}}{2} \right) \frac{\left( \kappa^2 + 4\nu^2 \right)\gamma_{\mathrm{l}} (2 n_{T} +1)   + k^2 \kappa (2 n_{\mathrm{s}}  +1)  }
    {[(4\nu^2+k^2) - \gamma\kappa ]^2/2 - \nu^2 (\gamma + \kappa)^2 } .
\end{align}
At the cavity resonance this becomes 
\begin{align}
    S(0) & = \left( \frac{\gamma_{\mathrm{c}}}{2} \right) \frac{\kappa^2 \gamma_{\mathrm{l}} (2 n_{T} +1)   + k^2 \kappa (2 n_{\mathrm{s}}  +1)  }
    {[k^2 - \gamma\kappa ]^2  } .
\end{align}

The noise added by the parametric amplifier at the input is   
\begin{align}
    \mathcal{N}_{\mathrm{add}} & = \frac{S(0)}{G_{\mathrm{ss}}^2} = \frac{\kappa^2 \gamma_{\mathrm{c}}\gamma_{\mathrm{l}} (2 n_{T} +1)   + k^2 \gamma_{\mathrm{c}}\kappa (2 n_{\mathrm{s}}  +1)  }
    {G_{\mathrm{ss}}^2 [k^2 - \gamma\kappa ]^2 } \\
    & = \frac{\gamma_{\mathrm{c}}}{\gamma}\frac{ (\gamma_{\mathrm{1}}/\gamma) (2 n_{T} +1)   + k^2/(\gamma\kappa)  (2 n_{\mathrm{s}}  +1)  }
    { [(\gamma_{\mathrm{c}} - \gamma_{\mathrm{l}})/\gamma +
    k^2/(\gamma\kappa) ]^2 }. \nonumber
\end{align} 
Note that since we must have $k < \!\sqrt{\gamma\kappa}$, we always have $k^2/(\gamma\kappa) < 1$. To minimize the added noise, we want to make $\gamma_{\mathrm{l}}$, which is the spin decoherence rate plus any internal cavity losses as small as possible compared to the input/output coupling $\gamma_{\mathrm{c}}$. 
 
Since the noise added at the input by an amplifier with noise temperature $T$ is   
\begin{align}
    \mathcal{N}_{T} & = n_{T} + \frac{1}{2},
\end{align} 
we write $\mathcal{N}_{\mathrm{add}}$ in the form 
\begin{align}
    \mathcal{N}_{\mathrm{add}} = n_{\mathrm{add}} + \frac{1}{2}
\end{align}
and obtain 
\begin{align}
    n_{\mathrm{add}} & = \left(\frac{\gamma_{\mathrm{c}}}{\gamma}\right) \frac{ (\gamma_{\mathrm{1}}/\gamma) (2 n_{T} +1)   + k^2/(\gamma\kappa)  (2 n_{\mathrm{s}}  +1)  }
    { [(\gamma_{\mathrm{c}} - \gamma_{\mathrm{l}})/\gamma +
    k^2/(\gamma\kappa) ]^2 }  -\frac{1}{2} .
\end{align}

\section{Efficacy of the Holstein-Primakoff approximation at 80\% polarization}
\label{HPcheck}

\begin{figure}[t]
    \centering
    \includegraphics[width=0.95\columnwidth]{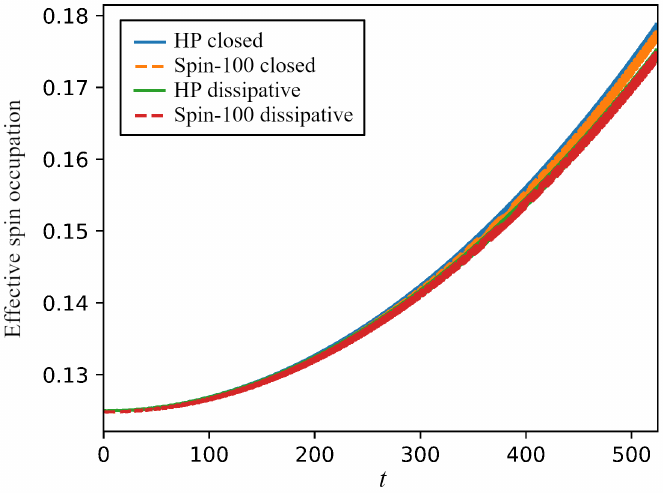}
    \caption{
    Comparing the Holstein-Primakoff approximation with a simulation using a spin-100 with $P=80\%$. The comparison is shown for both closed and dissipative evolution. The residual spin-sector offset is consistent with the leading finite-$J$ correction, while the maximum excitation fraction remains below $10^{-3}$.}
    \label{fig:hp_finite_spin_sanity}
\end{figure}

It is standard practice to assume that the linear approximation for spins works well at 80\% polarization, since experimental results in this regime show linear behavior. However, to check this approximation, we performed a numerical simulation using a single spin with $J=100$ to compare it against the Holstein--Primakoff approximation at $P=80\%$ (equivalently $n_s=(1-P)/(2P)=1/8$). The spin-100 simulation uses the effective operator
\begin{align}
    b_{\rm eff}=\frac{J_-}{\sqrt{2J}},
\end{align}
with the coupling rescaled as $g_{\rm single}=g_{\rm collective}/\sqrt{2J}$ so that the spin-100 Hamiltonian approaches the Holstein--Primakoff oscillator Hamiltonian in the low-excitation limit. For the dissipative comparison, the spin damping in the spin-100 dissipator was correspondingly scaled as $\kappa_{\rm single}=\kappa_{\rm HP}/(2J)$ because the Lindblad dissipator is quadratic in the annihilation operator.

The benchmark was performed in the Fock basis for the cavity mode with a photon-number cutoff of $N_{\rm cav}=20$, along with $P=0.8$, $\Lambda/(2\pi)=0.5~\mbox{GHz}$, $\omega_c/(2\pi)=2.5~\mbox{GHz}$, $\omega_s/(2\pi)=3.5~\mbox{GHz}$, and $g_{\rm collective}/(2\pi)=0.01~\mbox{GHz}$, for both closed and dissipative evolution. We show the results in Fig.~\ref{fig:hp_finite_spin_sanity}. The maximum excitation fraction remained below $10^{-3}$ in both cases. The cavity quadrature variances agree at the level of $10^{-4}$--$10^{-5}$, while the spin quadrature variances agree at the level of $10^{-3}$. The residual spin-sector offset is consistent with the leading finite-$J$ correction
\begin{align}
    J_- = \sqrt{2J}\sqrt{1-\frac{b^\dagger b}{2J}}\,b .
\end{align}

\bibliography{references}

\end{document}